# Structure and thermodynamics of supported lipid membranes on hydrophobic van der Waals surfaces

Harriet Read[a,b], Simone Benaglia[a,b], Laura Fumagalli[a,b]
[1]Department of Physics & Astronomy University of Manchester, Manchester, M13 9PL, UK
[2]National Graphene Institute, University of Manchester, Manchester, M13 9PL, UK

**Understanding the adsorption and physical characteristics of supported lipid membranes is crucial for their effective use as model cell membranes. Their morphological and thermodynamic properties at the nanoscale have traditionally been studied on hydrophilic substrates, such as mica and silicon oxide, which have proved to facilitate the reconstruction of biomembranes. However, in more recent years, with the advent of the van der Waals crystals technology, two-dimensional crystals such as graphene have been proposed as potential substrates in biosensing devices. Membranes formed on these crystals are expected to behave differently owing to their intrinsic hydrophobicity, however thus far knowledge of their morphological and thermodynamic properties is lacking. Here we present a comprehensive nanoscale analysis of the adsorption of phosphatidylcholine lipid monolayers on two of the most commonly used van der Waals crystals, graphite and hexagonal boron nitride. Both morphological and thermodynamic properties of the lipid membranes were investigated using temperature-controlled atomic force microscopy. Our experiments show that the lipids adsorb onto the crystals, forming monolayers with their orientation dependent upon their concentration. Furthermore, we found that the hydrophobicity of van der Waals crystals determines a strong increase in the transition temperature of the lipid monolayer compared to that observed on hydrophilic substrates. These results are important for understanding the properties of lipid membranes at solid surfaces and extending their use to novel drug delivery and biosensing devices made of van der Waals crystals.**

Supported lipid bilayers (SLB) are a fundamental and widely-used experimental platform in biophysics.[1,2] By integrating other biomolecular complexes, they allow for reconstruction of model cell membranes and investigation of their structural and physical-chemical properties *in vitro*. For example, they have been used to study transmembrane proteins,[3,4] and to visualise time-dependent processes such as protein-lipid and drug-lipid interactions[5,6], as well as molecular recognition.[7] Atomic force microscopy (AFM) based techniques have been extensively used to study SLBs for their ability to access morphological and physical properties of the lipid membranes down to atomic scale resolution and with precise control over the system's environment.[8,9] High-resolution AFM has also been deployed to resolve the morphological organisation of lipids together with their water solvation structures.[10,11] Additionally, AFM spectroscopy has been used consistently to determine the nanomechanical behaviour of supported lipid bilayers.[12–14]

SLBs spontaneously form through self-assembly on various flat substrates using either vesicle fusion methods[15,16] or the Langmuir-Blodgett technique.[16,17] Hydrophilic materials, such as glass, mica and silicon dioxide, have been commonly used as substrates as they facilitate the formation of stable lipid bilayers in water solutions, mimicking the structure of cell membranes in their native environment.[18–20]

The lipid molecules in the bilayer form two adjacent leaflets where the head groups expose towards the water and the substrate - known as the distal and proximal leaflets, respectively - whilst their hydrophobic tails are buried inside. Only recently has the use of hydrophobic carbon substrates such as graphite and its monolayer counterpart graphene been introduced as supports for the formation of lipid membranes.[21–26] In this case, lipids immersed in water form stable monolayers with the tails adjacent to the solid interface and the head groups pointing away from the hydrophobic surface towards the water. It has been predicted that subsequent lipid bilayers might then form on top of the first interfacial monolayer.[23] Being electrically conductive,[27] graphite/graphene substrates are an ideal platform to carry out bio-electrochemical experiments on lipid membranes[28], often used as a substitution for gold substrates which have been traditionally employed for biosensing applications.[29,30] Hence, with the recent advances in nanoscience and two-dimensional (2D) van der Waals (vdW) technology that allow the development of novel 2D nano-sensors, it has become crucial to prepare, in a controlled and reproducible manner, lipid membranes on graphene and other vdW crystals.

A key feature of lipid membranes is their ability to exist in different thermodynamic phases, which crucially depends upon their chemical composition and environmental conditions. Additionally they also undergo reversible phase transitions, as characterised by their phase transition temperatures.[31,32] In particular, the melting temperature, $T_m$, indicates the main phase transition of lipids from a solid-ordered ($S_o$) phase, where the lipids are regularly packed with their tails extended, to a liquid-disordered ($L_d$) phase, where the lipids' tails compress and the lipids are more free to diffuse laterally, leading to a less ordered membrane. This results in an overall 'shrinking' of the lipid membrane thickness. Depending upon the model system the phase transition can be studied by numerous techniques. Commonly, differential scanning calorimetry (DSC) is used as the gold-standard technique for multi- or uni-lamellar vesicles in solution, whilst the phase transitions of SLBs have been studied using spectroscopic techniques such as Raman spectroscopy,[33,34] sum frequency generation spectroscopy,[33,35] and AFM.[18,36–44] Importantly, SLBs have demonstrated different behaviour compared with lipid vesicles due to the presence of the solid support.[37,42] Only recently has a clear understanding of the thermodynamic behaviour of the two lipid leaflets constituting the bilayer been achieved.[35] Whilst the phase transition of vesicles occurs at a single temperature, two transitions have been observed for SLBs on hydrophilic surfaces such as mica and silicon. The first occurs at a temperature similar to that found for vesicles and corresponds to the phase transition of the distal leaflet. The second occurs at a higher temperature and is associated to the melting of the proximal leaflet - sitting adjacent to the solid surface. This decoupled effect is a consequence of the strong interaction between the hydrophilic substrate and the polar heads of the proximal lipid monolayer,[18,36–38,41] inducing a different lipid density in the two leaflets.[45–47] Interestingly, it has been suggested that by modulating the environmental and preparation conditions of SLBs, one may couple or decouple the phase transition of the two leaflets.[38] However, to the best of our knowledge, no studies have been reported on the effect of hydrophobic substrates on the thermodynamics of supported lipids membranes.

Here, we utilised temperature-controlled amplitude modulation AFM to investigate the morphological and thermodynamic properties of two commonly used phosphatidylcholine (PC) lipids (1,2-dimyristoyl-sn-glycero-3-phosphocholine, DMPC, and 1,2-dilauroyl-sn-glycero-3-phosphocholine, DLPC) deposited, via vesicle-fusion methods, on two hydrophobic vdW crystals: highly ordered pyrolytic graphite (HOPG) and hexagonal boron nitride (h-BN). First, the adsorption and structural arrangement of lipids on HOPG was investigated. Secondly, we studied the effect of the temperature on fully formed lipid monolayers and subsequently determined the transition temperature for both DMPC and DLPC monolayers on HOPG. We studied the nanoscale morphological structure of lipid monolayers below and above the transition temperature. We contrasted these results with those obtained on hydrophilic supports by repeating the experiments on mica and silicon oxide substrates, ultimately finding important differences in the transition temperature. To understand the impact of graphite conductivity on these findings, we analysed the case of h-BN, a similar vdW crystal to HOPG as it shares the same hydrophobic character and hexagonal lattice structure but importantly differs in its electrical properties with it being electrically insulating.[48] We found a large increase of the transition temperature of PC lipids on HOPG with respect to those obtained on lipids membranes on hydrophilic substrates and on h-BN, indicating the important role of the substrate metallicity on the structure and phase of lipid membranes.

## Materials and Methods

### Lipid sample preparation

Multi-lamellar liposomes were attained following methods reported by Atwood et al.,[15] in order to prepare samples via vesicle fusion for analysis using AFM. PC lipids, DMPC and DLPC, were bought from Avanti Lipids and stored at -20 °C. Chloroform (anhydrous, ≥99%, Sigma-Aldrich) was added to both to make stock solutions of 10 mg/ml, which were stored in amber vials to reduce oxidation. Under a nitrogen flow, the chloroform was evaporated whilst rotating the vial to form an even film. The vial was left overnight to ensure the absence of any chloroform. The lipid film was re-hydrated using deionised (DI) water, of resistivity 18.2 MΩ (Millipore), to form concentrated lipid suspensions and was followed by 10 minutes of sonication to remove any leftover film from the vial. No further extrusion was carried out, however before producing the diluted suspensions used for deposition, the concentrated suspension was sonicated for 10 more minutes. The liposome suspensions were stored away from light, at approximately 4-8 °C.

For experiments on hydrophilic surfaces, mica and p-doped silicon (with its native oxide), 0.2 mg/ml concentration suspensions of DMPC-water were deposited on the surface and left to incubate at room temperature for 10 minutes before rinsing with DI water, ensuring a water droplet remained on the surface. Prior to deposition, mica surfaces were freshly cleaved, and silicon chips were treated with a piranha solution (9:1 sulphuric acid:hydrogen peroxide, Sigma-Aldrich) at 80 °C for two minutes before rinsing with DI water. Silicon chips were then sonicated with a solution (5% in DI water) of Decon-90 (Decon Laboratories Ltd, UK). For hydrophobic surfaces, DMPC and DLPC suspensions of concentrations ranging from 1 µg/ml to 0.1 mg/ml were used. Around 200 µl of the suspension was dropped onto freshly cleaved HOPG/hBN and was left to incubate for 30 minutes at 40 °C before rinsing with water, as with the hydrophilic surfaces. For both types of surfaces, the samples were immediately taken to the AFM for measurements ensuring that the sample remained hydrated.

### AFM measurements

AFM measurements were carried out in water solutions using a commercial AFM (Cypher ES, Asylum Research, Oxford Instruments) in amplitude modulation with photothermal excitation of the cantilever. Gold-coated cantilevers (HQ NSC19/Cr-Au, µMasch) were used, with a typical spring constant of ~ 1 N/m and resonance frequency of ~ 35 kHz in liquid. The temperature-controlled sample stage allowed us to investigate the temperature dependence of the supported lipid membranes between 15 °C and 70 °C, with a heating rate of 0.1 °C/s. The sample was left for 2-5 minutes to thermalise before re-approaching the AFM tip to the surface and resuming imaging. Care was taken to re-align the scan at each temperature to ensure the same areas were imaged consecutively. Furthermore, multiple scans at each temperature were performed to ensure that the sample had thermalised. Post processing and analysis of the AFM images were done using WSxM and Gwyddion software. To image the lipid ripple structures described below, very small oscillation amplitudes (< 1 nm) were applied to the AFM cantilever, as typically done to obtain high-resolution images of different materials in both air and liquid environment.[49–52] Cantilevers were calibrated by taking force-distance curves towards the sample and the static and first mode optical lever sensitivity, $\sigma_0$ and $\sigma_1$ ($\sigma_1 = 1.09\sigma_0$), were extracted from the slope of the deflection (V)-piezo extension (nm) curve.

## Results and discussion

### Growth and structure of DMPC lipid membranes

We started by investigating the formation of DMPC lipid membranes on HOPG. Initially, low concentrations (1 µg/ml) of DMPC/DI suspensions were deposited on freshly cleaved HOPG surfaces, and time-lapse AFM images were taken to visualise the growth of the lipid membrane. For such low concentrations, we found the presence of small patches at the beginning of the imaging that, under stable conditions, continuously grew up until a saturation point. Figure 1a shows three consecutive images of a representative patch taken at times t = 0, 6 and 14 minutes, showing the structural evolution of the patch. The area covered by the patch increases with time due to the continued adsorption of lipid molecules from the lipid suspension. Corresponding topography profiles (Figure 1b) reveal that the height of the lipid patch does not change with time and is ~ 0.6 nm. This value is much lower than expected for a DMPC monolayer (~ 2.2 nm)[29,30], indicating that at such low concentrations the lipid molecules lie essentially parallel to the HOPG surface. We found this configuration only for dilute vesicle suspensions which yielded small monolayer patches on the surface. Such a tilted, flat-lying, conformation has been previously reported for DMPC deposited on hydrophobic gold[29,30] and is also typically found for alkanes molecules on HOPG.[53] As for the case of alkane molecules, this orientation of DMPC molecules can be explained by the high-affinity of the lipid alkyl chains for the HOPG surface.[54]

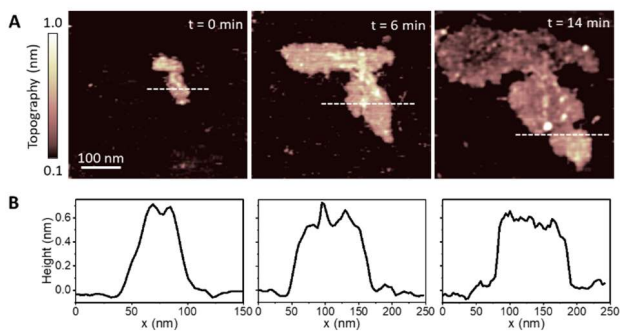

**Figure 1.** (a) AFM topography images of the adsorption of DMPC molecules on a HOPG substrate. Images were taken at consecutive times (0, 6, and 14 min) in a water after depositing a droplet of DMPC/DI suspension at low concentration (~ 1 μg/ml). (b) Corresponding height profiles taken across the dotted line in the topography images. They show the height of the lipid patch to be around 0.6 nm and constant with time, indicating a tilted flat-lying orientation of the lipid molecules.

Next, we increased the incubation concentration of DMPC vesicles. Using approximately two orders of magnitude higher concentration (~ 0.1 mg/ml), a full-supported lipid membrane on HOPG was formed, with few defects to reveal the bare HOPG substrate below. These defects allowed us to measure the thickness of lipid membrane with respect to the HOPG. Figure 2a shows a representative topography image of the fully formed DMPC monolayer on HOPG.

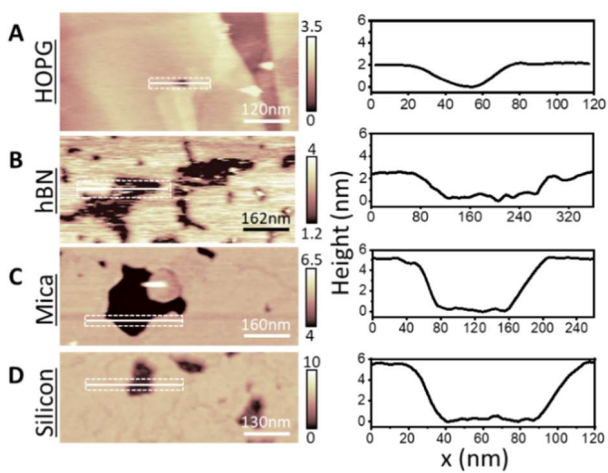

**Figure 2.** From top to bottom: DMPC lipid membranes formed on (a) HOPG, (b) hBN, (c) muscovite mica, and (d) on silicon at temperature < 20 °C and corresponding cross sections. A DMPC monolayer with thickness ~ 2 nm was formed on HOPG and hBN, whilst a bilayer was obtained on mica and silicon with thickness ~ 5 nm. The height of the monolayer on the hydrophobic surfaces indicates that the lipid molecules are oriented perpendicularly to the surface with their tails towards the substrate.

The cross section taken across a defect shows that the thickness of the formed layer is 1.9 ± 0.1 nm, which was furthermore confirmed by performing force-distance curves upon the layer, yielding 2.2 ± 0.2 nm (see Figure S5 in ESI). The obtained thickness matches the value expected for the DMPC monolayer, indicating that for membranes formed from higher concentration suspensions, the DMPC molecules are oriented perpendicularly to the HOPG with the tails adjacent to the substrate. Again, this is consistent with results previously obtained on hydrophobic gold.[29,30] Importantly, these measurements were performed at a temperature < 20 °C to ensure that the lipid layer was in its expected $S_o$ phase. As a control of our experimental method, we carefully repeated the DMPC deposition on hydrophilic surfaces of muscovite mica and silicon, which have been previously used as supports of lipid bilayers.[38] In both cases, the AFM images (Fig. 2c and Fig. 2d) clearly indicate the formation of a bilayer of thickness ~ 5 nm in good agreement with expectation and previous reports for DMPC bilayers (average height of 5.3 ± 0.5 nm and 5.0 ± 0.2 nm for mica and silicon respectively).[55]

The thickness of the formed lipid membrane was not the only morphological difference between the case of HOPG and that of the hydrophilic surfaces. AFM images taken at higher resolution[50–52] (ESI Fig. S4) revealed the existence of stripe-like domains that extended across the whole DMPC lipid monolayer surface, unlike the smooth surface seen for hydrophilic supports. To investigate this, we obtained images at various temperatures ranging from 21 °C up to 60 °C. The stripe-like domains, which displayed an average periodicity of ~ 8 nm, were stable and did not change significantly with temperature. This rules out that these features are a result of the so-called ripple phase seen in SLBs, associated with alternating domains at temperatures near the main phase transition of the DMPC bilayer.[39,41,56] Instead, they are consistent with previous reports of stripe-like domains for lipids deposited on hydrophobic gold and HOPG supports,[24,29,30,57] suggesting that on hydrophobic materials the lipid monolayer organises into hemimicellar structures.

To better understand the impact of the surface hydrophobicity on the formation of PC lipid membranes, we investigated the behaviour of DMPC lipids on h-BN crystals, offering an alternative substrate to HOPG. h-BN is a similar van der Waals crystal to HOPG, sharing features such as moderate hydrophobicity and atomic flatness. Furthermore, h-BN is structurally a-like graphite, displaying a honeycomb atomic lattice, however possessing a boron–nitride pair instead of the double bonded carbon atoms, as found for HOPG. This results in h-BN being electrically insulating, as opposed to HOPG which is a semi-metal, and thus the two surfaces display very different conductive properties. As for HOPG, DMPC lipid membranes were formed on h-BN via vesicle fusion via the same procedure. We obtained similar monolayer formation, as for the case of HOPG (see Fig.2b). Importantly, we note the existence of many more defects in the lipid monolayer on h-BN that revealed the bare h-BN surface, in contrast to the few present on HOPG. As before, from the defects we could directly measure the layer thickness (Fig.2b). AFM cross sections yielded a membrane thickness of ~ 2.2 nm, in agreement with values found on HOPG and again indicating the formation of a lipid monolayer.

**Temperature-dependent behaviour of DMPC monolayers**

Next we analysed the temperature-dependent behaviour of DMPC membranes on hydrophobic and hydrophilic surfaces using the established van't Hoff analysis. As previously mentioned, a change in the lipid phase can be recognised by the different height of lipids domains in the $S_o$ and $L_d$ phases.[44] The main phase transition temperature, $T_m$, can then be obtained by fitting the fraction of $S_o/L_d$ domains (as a function of temperature) with a sigmoidal function, i.e. using the van't Hoff equation.[32] Writing $s$ and $l$ as the fractional occupancy of the $S_o$ and $L_d$ domains, respectively, we can define the equilibrium constant, $K = \frac{s}{l}$, of the phase transition in terms of the temperature as

$$lnK = \frac{\Delta H_{vH}}{R}\left(\frac{1}{T_m} - \frac{1}{T}\right) \quad (1)$$

where $\Delta H_{vH}$ and $R$ are the van't Hoff enthalpy and the gas constant, respectively. If we express $s$ as a function of $T$, we then obtain

$$s = \frac{1}{1+\exp\left(\frac{\Delta H_{vH}}{R}\left(\frac{1}{T_m} - \frac{1}{T}\right)\right)} \quad (2)$$

which is a sigmoidal function that can be used to describe the behaviour of the fractional occupancy of the $S_o$ lipid domains.[38,41] First, we verified this procedure on DMPC bilayers on hydrophilic surfaces, analysing the changes in the AFM images taken at increasing temperature. DMPC lipids have an expected transition temperature of approximately 23 °C, as determined from liposome suspensions by DSC experiments.[58] Figure 3c shows the fractional occupancy of the lipid bilayers in the $S_o$ phase as a function of temperature as obtained on mica (green symbols) and silicon (blue symbols) substrates (see also Fig. S1 and Fig. S2 in ESI). We found similar behaviour for both substrates, showing two transition temperatures corresponding to a decoupled phase transition of the distal and proximal leaflets, with the transition of the proximal leaflet occurring at a higher temperature, in agreement with previous results for lipid bilayers on hydrophilic substrates.[18,36–38,41,59] Specifically, we found the $T_m$ of the distal and proximal layer to be 22.2 ± 0.1 °C and 33.8 ± 0.1 °C on mica, and 19.5 ± 0.1 °C and 30.7 ± 0.1 °C on silicon, respectively. Notably, for both substrates, the transition temperature of the distal layer is consistent with those found in DSC experiments (~ 23 °C),[58] as expected.

Next, we looked at the temperature-dependent properties of the DMPC monolayer on HOPG. As for the hydrophilic surfaces, we performed AFM topography images at increasing temperatures. In contrast with the previous experiments on hydrophilic substrates, we did not observe the appearance of domains with changing temperature but rather an overall change in the thickness of the monolayer as measured via the defects. From this, we were able to determine the transition temperature (see ESI S2 section for further details). Given that only a DMPC monolayer is present on the HOPG substrate, a temperature dependence similar to the proximal layer of DMPC bilayers on hydrophilic substrates was expected, i.e. a phase transition temperature occurring at higher temperatures than found in DSC measurements. Indeed, we found that the transition did not occur up to approximately 40 °C. Figure 3a shows topography images and height profiles at temperatures 40 °C, 50 °C and 60 °C. At 40 °C and 50 °C, the average thickness of the monolayer was found to be 1.9 ± 0.1 nm and 1.8 ± 0.1 nm, respectively, remaining practically constant to the initial thickness measured at 20 °C. Upon further increasing the temperature to 60 °C, the membrane thickness decreased to 1.4 ± 0.1 nm, which is close to the value reported for a DMPC monolayer in the $L_d$ state.[60] This implies the phase change occurs at around 50 °C, much higher than expected. Figure 3c shows the observed height of the lipid monolayer as a function of the temperature (red symbols) and the fitting with a sigmoidal function (red solid line), which yields $T_m$ = 52.6 ± 0.1 °C. This is approximately 20 °C higher than the value we found for the proximal leaflet on mica and silicon and 30 °C higher than that found for the distal leaflet.

As for the morphology, we compared the behaviour of the DMPC monolayer on HOPG with that on h-BN by repeating the temperature-dependent experiments (Figure 3b). Again, from the formation of domains with higher/lower height, we could recover the fractional occupancy of the lipid $S_o$ phase (as shown for the SLBs on hydrophilic substrates). Surprisingly, the transition temperature on h-BN is clearly smaller than the one obtained for HOPG (Fig. 3c, pink symbols). By fitting the data to Eq. 2, we found $T_m = 36.7 ± 0.1$ °C, which is close to the value obtained for the $T_m$ of DMPC proximal leaflet on hydrophilic materials.

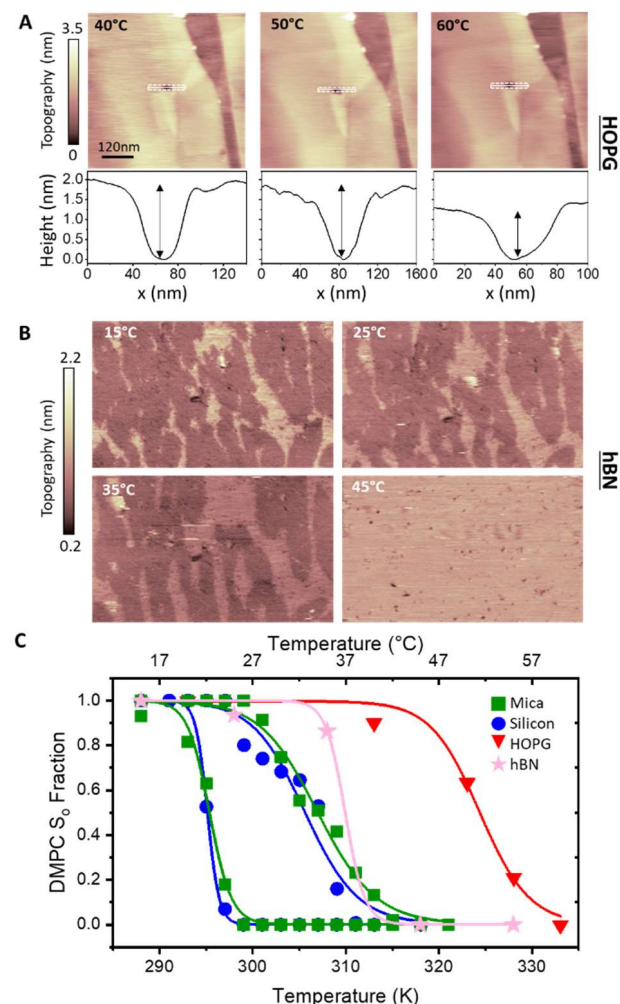

**Figure 3.** (a) Series of AFM images in water of a DMPC monolayer on HOPG (concentration 0.1mg/ml) at temperatures 40°C, 50°C, and 60°C, respectively, and corresponding height profiles taken along a defect in the monolayer indicated by the dashed lines. (b) Same as (a), but on h-BN at different indicated temperatures. (c) The experimental fractional occupancy of the lipid solid phase of DMPC monolayers on HOPG (red triangles) and h-BN (pink stars). For comparison, control data taken on DMPC bilayers on hydrophilic surfaces of mica (green squares) and silicon (blue circles) are also reported, showing both the distal and proximal leaflet transition. Solid lines are fittings to Eq. 2.

**Temperature-dependent behaviour of DLPC monolayers**

To verify whether the anomalous transition temperatures observed on hydrophobic surfaces for DMPC also occurs for other PC lipids, we proceeded to repeat the previous experiments with DLPC. DLPC was chosen as its expected $T_m$ from $S_o$ to $L_d$ phase is at approximately -1 °C.[58] This is much lower than the $T_m$ for DMPC and therefore helps to avoid experimental difficulties at high temperatures, in particular the evaporation of the water solution during the experiment. Following the same protocol as for DMPC, full DLPC membranes were formed on HOPG and h-BN. The thickness of these layers, determined via AFM topographic images and static force curves (see Fig. S5, S6, S7 in ESI), was found to be ~ 1.1 nm, on both HOPG and h-BN indicating the formation of a lipid monolayer. As for the case of DMPC, we found that DLPC monolayers presented stripe-like nano-domains on both HOPG and hBN with a periodicity of ~ 5 nm (see ESI Fig S4 and S7).

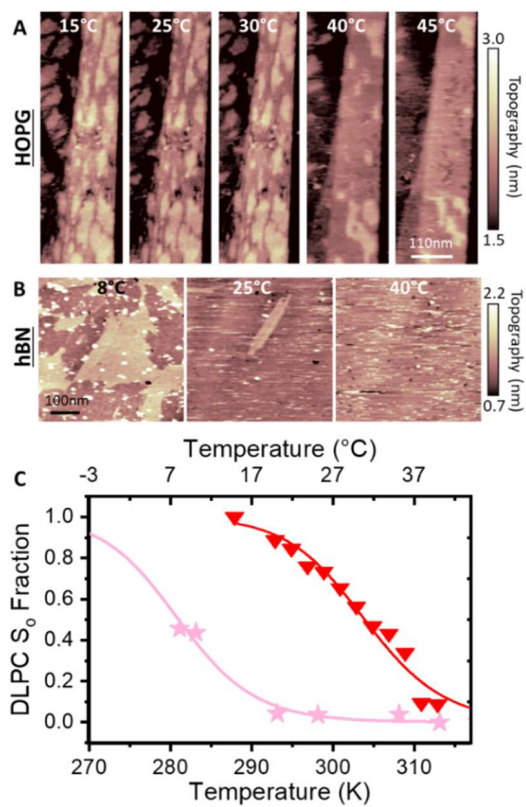

**Figure 4.** (a) AFM images in water of DLPC monolayer on HOPG (concentration 0.1mg/ml) at various indicated temperatures. Domains with a larger thickness appear as the lipid monolayer changes from $L_d$ to $S_o$ phase. As with DMPC, the transition of DLPC on HOPG is much higher than that reported with DSC (~-1°C). (b) Same as (a) but on h-BN. (c) Fractional occupancy of the solid phase as a function of the temperature on HOPG (red triangles) and h-BN (pink stars). Solid lines are fittings of Eq. 2.

AFM images of the DLPC monolayer on HOPG were recorded as a function of the temperature between 15 °C and 45 °C. Figure 4a shows that at 15 °C, the lipid monolayer was characterised by the high density of domains approximately 0.3 nm higher than the adjacent lipids (also shown is a terrace of the graphite substrate). Such domains correspond to the $S_o$ phase and were found over the whole surface. The fraction of the membrane consisting of higher domains changes with temperature, as it decreases at higher temperatures. At 45 °C they almost disappeared, indicating that the monolayer transitioned to its $L_d$ phase. The temperature was cycled, cooling and heating the sample multiple times whilst taking AFM images, showing the reversibility of the process (see the full temperature cycle in Fig. S8 in ESI), as previously shown for lipid bilayers on hydrophilic substrates.[37,43] Moreover, analysis for both heating and cooling yielded very similar results. By plotting the fractional occupancy of the $S_o$ domains with respect to the total area against the temperature and the fitting the data to Eq. 2 (Fig. 4c, red), we obtained the transition temperature of DLPC to be $T_m = 30.9$ °C, around 30 °C higher than expected from DSC.[58] This confirmed our previous results obtained on DMPC and suggests that the transition temperature of PC lipids on HOPG substrates is shifted upwards by approximately 20-30 °C. It is also interesting to note that, as for DMPC, similar ripple structures were found on both the $L_d$ and $S_o$ phase (see ESI, Figure S4), again suggesting these stripe-like domains are not the ones commonly associated with the phase transition of lipids but rather a morphological feature characterising lipids on these substrates.

Next, we studied the temperature-dependent behaviour of DLPC on h-BN. Assuming a similar behaviour as on HOPG, the temperature was decreased down starting from 40° C. However, no significant structural changes were observed down to 20 °C. At 10 °C, domains of a higher thickness in the $S_o$ phase were observed approximately occupying half of the imaged area. Due to experimental limitations in our setup, temperature-dependent AFM measurements further below 8 °C were not possible. Despite this, we could proceed with extracting the transition temperature. Figure 4c shows the fraction of the $S_o$ phase with respect to the total area as a function of the temperature and the fitting to Eq. 2, yielding $T_m = 8.2$ °C. This value is higher than the expected transition from DSC, however, it is more than 20 °C lower than that observed on HOPG.

A summary of all the transition temperatures of DLPC and DMPC obtained in this work is shown in Table 1, together with previously reported values obtained from DSC.[61]

| | DSC | Mica Distal | Mica Proximal | Silicon Distal | Silicon Proximal | HOPG | hBN |
|---|---|---|---|---|---|---|---|
| **DMPC** | | | | | | | |
| $T_m$ (°C) | 23 | 22.2 | 33.8 | 19.5 | 30.7 | 52.6 | 36.7 |
| **DLPC** | | | | | | | |
| $T_m$ (°C) | -2 | --- | --- | --- | --- | 30.9 | 8.2 |

**Table 1** Summary of the thermodynamic properties of supported DMPC and DLPC lipid monolayers on HOPG and h-BN. Control data obtained on lipids bilayers on muscovite mica and silicon surfaces are also reported.[58]

## Discussion and conclusions

In this work, we used temperature-controlled AFM to study the effect of two hydrophobic vdW crystals, HOPG and h-BN, on the morphological and thermodynamic properties of supported PC lipid membranes. We found that, on these substrates, lipid molecules organised themselves into monolayers, whilst on the hydrophilic substrates a lipid bilayer formed. The self-assembly of lipid molecules into a monolayer is likely to be driven by the strong interaction that hydrophobic vdW materials have with lipid alkyl tails. When using low lipid concentrations, the lipids seemingly adsorbed with their long axis parallel to the surface. However, upon increasing the lipid concentration, we obtained a rearrangement of the lipid layers, with the lipid molecules stacked perpendicular to the materials surface. In the latter case, the probed lipid thickness matched the predicted thickness for a single monolayer, confirming previous structural measurements.[55] Moreover, at these higher concentrations, temperature-independent stripe-like domains were present. We disassociate this from the commonly observed 'ripple-phase' of SLBs which is known to be due to the temperature-dependent competing phases of the lipid membrane.[39] Rather, we argue that it is linked to the presence of the hydrophobic substrate, as already suggested in previous literature.[24,28,29]

We then characterised the phase transition of the lipid monolayers as a function of temperature. A general increase of the transition temperature with respect that found by DSC experiments was expected due to the interaction between the substrate and the lipid membrane. This has previously been reported for the proximal layer of the lipid bilayers on hydrophilic surfaces. Indeed, in our control experiments on mica and silicon supports we observed a decoupled phase transition of the distal and proximal leaflets, with the proximal leaflet showing higher transition temperatures. However, on HOPG we found that the phase transition shifted substantially by over 30 °C more than expected for both DMPC and DLPC. Previous reports have shown that the melting temperature of monolayers of linear alkyl molecules formed on graphite, which is qualitatively similar to the main phase transition of lipids from their $S_o$ to $L_d$ phase,[62] is expected to shift to higher temperatures with respect to the bulk.[63,64] Although this seems analogous to our experimental observation, to our knowledge, this effect has never been reported for lipid membranes.

Our experiments on h-BN, which is both hydrophobic and atomically flat similarly to HOPG, shows that the transition temperature of the DMPC lipid membrane shifts again to higher temperatures. However, by only a few degrees more than the proximal leaflet of DMPC SLBs. Additionally, repeating the experiments on h-BN with the DLPC lipid, we found a transition temperature more than 20 °C less than that found for HOPG, confirming that PC lipids on h-BN have quite a different thermodynamic behaviour with respect to HOPG. To rationalise our observations, we note that in our experiments we see a reduction of defects in the monolayers on hydrophobic surfaces in comparison to SLBs on hydrophilic surfaces, suggesting a more regular lipid packing on hydrophobic surfaces. This is shown in Figure 5. Additionally, it is apparent that the frequency of defects in lipid monolayers on HOPG in comparison to h-BN is reduced.

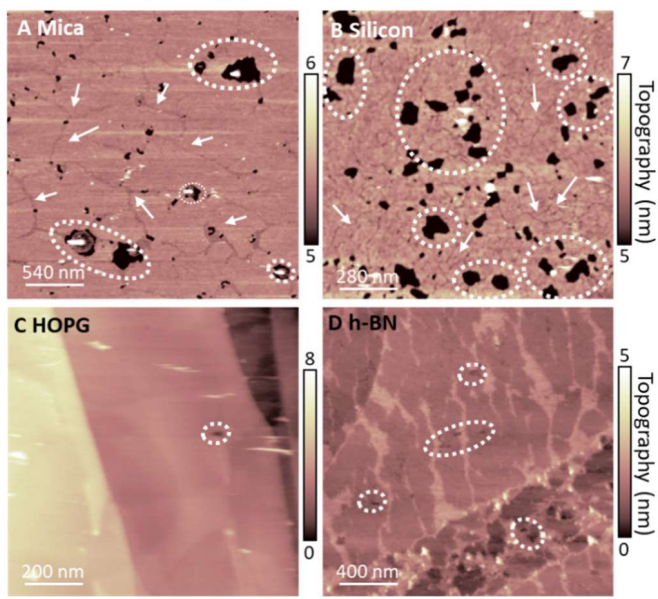

**Figure 5.** DMPC lipid layers on hydrophilic (mica and silicon) and hydrophobic surfaces (HOPG and hBN) in their $S_o$ phase, showing presence of defects in the membranes. Examples of holes and cracks in the membrane are indicated by white circles and white arrows respectively (note that cracks are only present in the SLBs, which are a result of the packing irregularities of lipid molecules in SLBs). For hydrophilic surfaces, we noticed the presence of many holes and cracks. For the hydrophobic vdW surfaces we observed fewer defects, and we did not record cracks in the layer.

This is particularly relevant, since the main phase transition of lipid membranes is known to start from defects present in the membrane (so called 'cracks' and holes), which arise due to packing irregularities of the lipid molecules.[46,47] This is concurrent with our results, presented in ESI Fig. S1,2, where the transition can be seen to originate from the defects in the bilayer. Furthermore, higher density packing of lipids in the membranes has demonstrated an increase in the transition temperature of the membrane, as for the case of the proximal leaflet on hydrophilic substrates.[38,59] This can be further illustrated by the lower transition temperature of the SLB on silicon in comparison to mica, relating an increased frequency of membrane defects of the DMPC SLB on the silicon surface, as shown in Fig. 5. Hence, we argue that on hydrophobic supports (i) a highly packed lipid layer and (ii) the absence of defects in the monolayer lead to an increase in $T_m$. In addition, we observed that an even more highly ordered monolayer formed on HOPG than h-BN shifts the $T_m$ to even higher temperatures. Further investigation is needed to understand the origin of this effect. However, as h-BN and HOPG surfaces share similar honey-comb atomic lattice and only differ in their conductivity and surface charge, we speculate that this effect may originate from long-range forces, such as vdW forces, rather than steric constraints between the crystal lattice and the lipid alkyl chains.

In conclusion, we have studied the impact of hydrophobic vdW substrates on both the physical and thermodynamic properties of reconstructed lipid membranes. Our findings improve our understanding of lipid membranes' properties at solid surfaces and will be useful in various applications of lipid membranes, in particular the development of novel bioelectric devices as well as experimental biosensing setups using vdW crystals as electrodes or supports.

**Author Contributions**

L.F. conceived and supervised the project. H.R. and S.B. performed sample preparation, measurements and data analysis. H.R. and S.B. wrote the paper with contributions of L.F.

**Conflicts of interest**

The authors declare no competing interests.

**Acknowledgements**

The authors thank Lorena Redondo-Morata for fruitful discussions. This work received funding from the European Research Council (grant 819417 - Liquid2DM) under the European Union Horizon 2020 research and innovation programme and from the UKRI (grant EP/X022471/1 – ElectroProtein).

Supplementary Information

**S1 Phase transition of DMPC SLBs on mica and silicon**

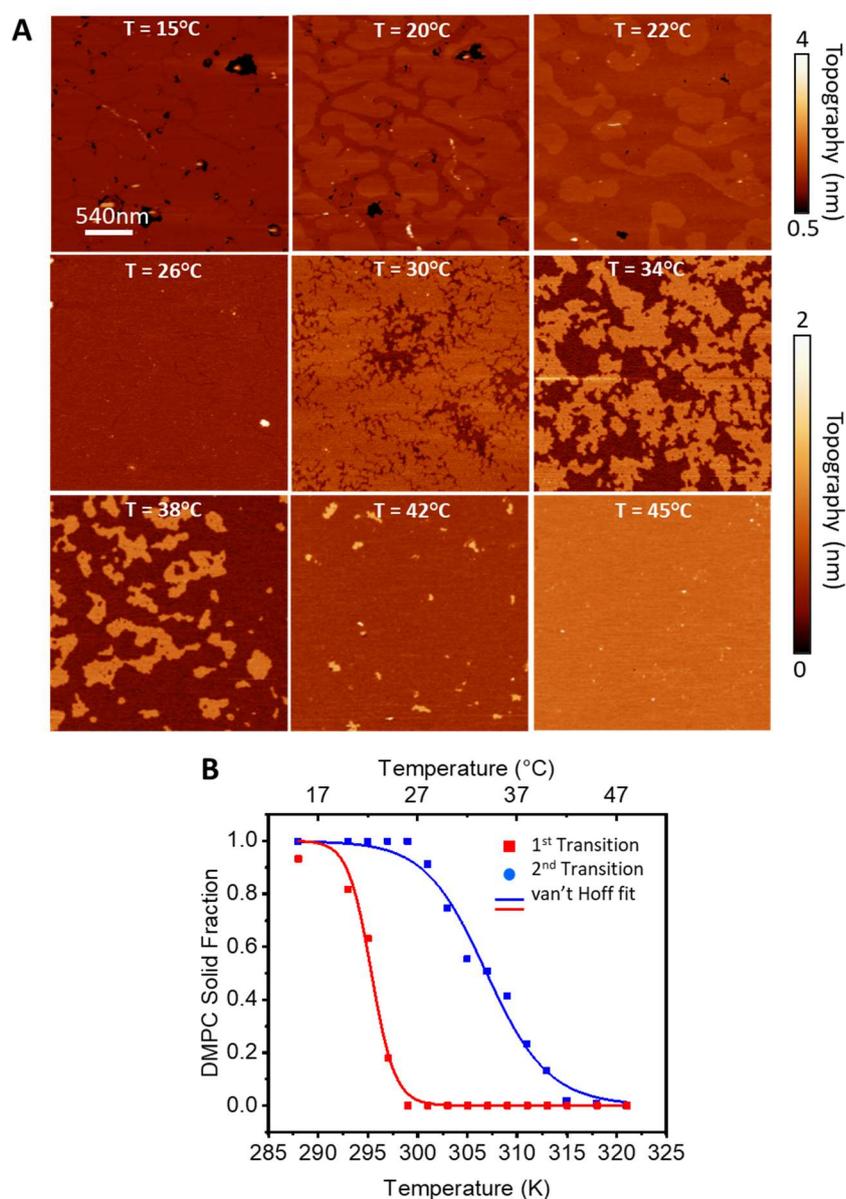

**Figure S1.** DMPC vesicles were formed in pure water following the same protocol in the main text and previous literature.[1] A complete DMPC supported lipid bilayer with small defects formed on freshly cleaved mica, using 0.2 mg/ml DMPC/DI suspension. (a) The temperature-dependent phase transition of the bilayer was tracked using AFM. Increasing the temperature led to a decoupled leaflet transition, with the top leaflet (lipid layer facing the aqueous solution) transitioning from the gel ordered to fluid disordered phase. Once this transition was completed, the second, lower leaflet (adjacent to the substrate) made the transition. The temperature was increased until all parts of the bilayer were transitioned. (b) To analyse the transition temperature of each leaflet separately, the fraction of the gel phase vs temperature was plotted and fitted using van't Hoff analysis. This revealed a transition temperature of the upper/lower leaflet as 22 °C (295 K) and 34 °C (307 K), respectively. The second transition is much higher than the expected transition temperature (determined using scanning differential calorimetry) due to the interaction between the substrate and the lipids.[2]

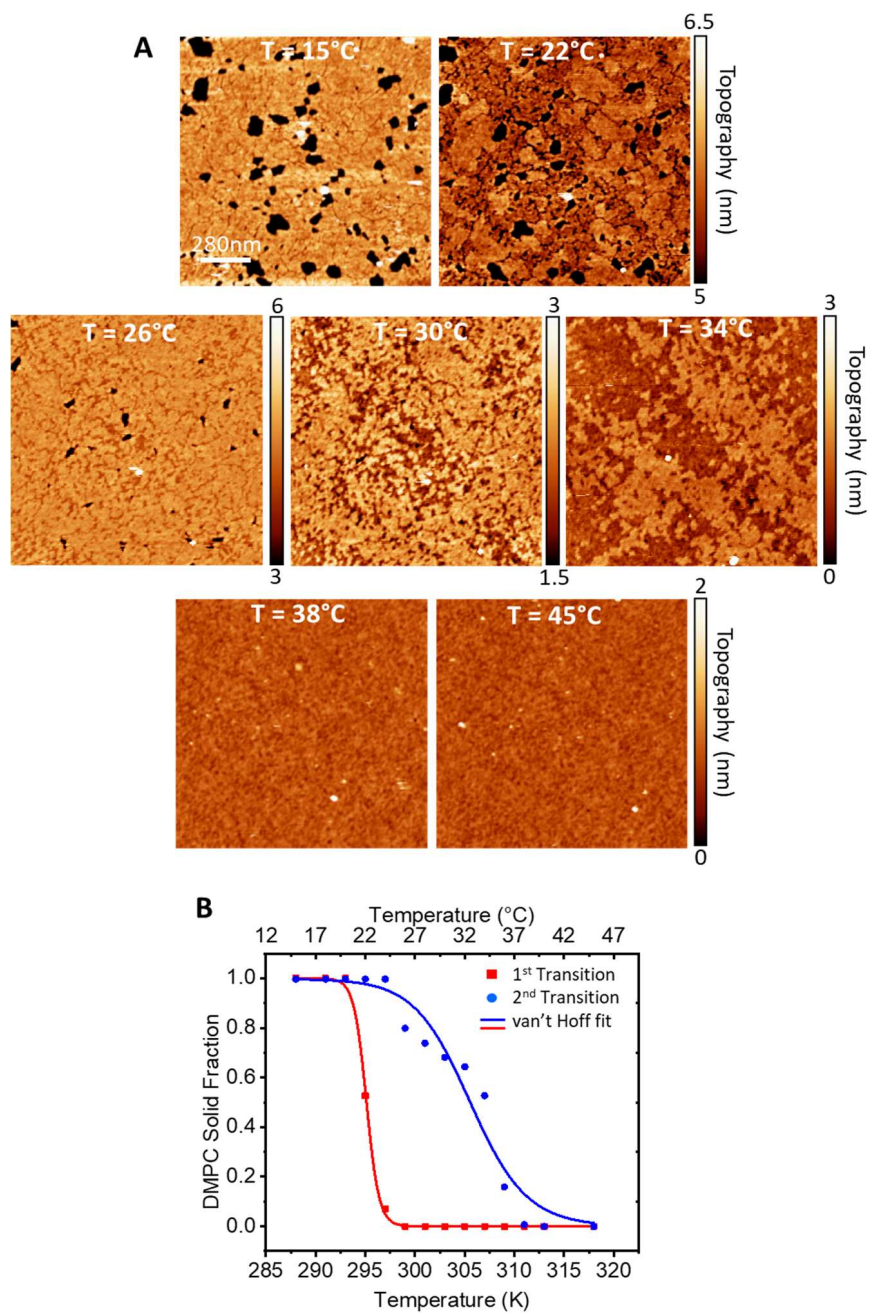

**Figure S2.** A complete DMPC supported lipid bilayer with small defects formed on clean silicon wafers using 0.2 mg/ml DMPC/DI suspension. (a) The temperature-dependent phase transition of the bilayer was tracked using AFM in liquid environment. Increasing the temperature led to a decoupled leaflet transition, with the top leaflet (lipid layer facing the aqueous solution) transitioning from the gel ordered to fluid disordered phase, as was seen on mica. (b) To analyse the transition temperature of each leaflet separately, the fraction of the gel phase vs temperature was plotted and fitted using van't Hoff analysis. This revealed a transition temperature of the upper/lower leaflet as 22 °C (295 K) and 32 °C (307 K), respectively. The transition temperature of the first leaflet is very close to the one extracted on mica and those found via DSC. The second transition is higher than the expected transition temperature due to the interaction between the substrate and the lipids.

**S2 Phase transition of DMPC monolayer on HOPG**

To compare the results obtained on HOPG with those on mica and silicon, the fraction of the $S_o$/$L_d$ phase was estimated from the monolayer height change on HOPG (Figure 3a). The distributions in Figure S3b, as the cross sections in Figure 3 of the main text, show that the average height of the monolayer tends to decrease upon increasing the temperature. The change in height with respect to the temperature is plotted in Figure 3c. To convert the change in height into fraction of the $S_o$, we assumed that the average height of the membrane at a temperature T, H(T), is given by the sum of height in the $S_o$, $h_S$, and $L_d$, $h_L$, phase normalised by the fraction of the membrane in the $S_o$, $f_S$, and $L_d$, $f_L$, phase respectively such that: $H(T) = h_S f_S + h_L f_L$. Considering that the sum of the fraction of the membrane in the $S_o$ and $L_d$ phase may be written as $f_S + f_L = 1$, by simply rearranging for $f_S$ we obtain the following expression: $f_S = (H(T) - h_L)/(h_S - h_L)$. This allows a plot of the $S_o$ fraction vs temperature as we did for the mica and silicon substrates. The curve can be seen in Figure S3c and in Figure 3 of the main text.

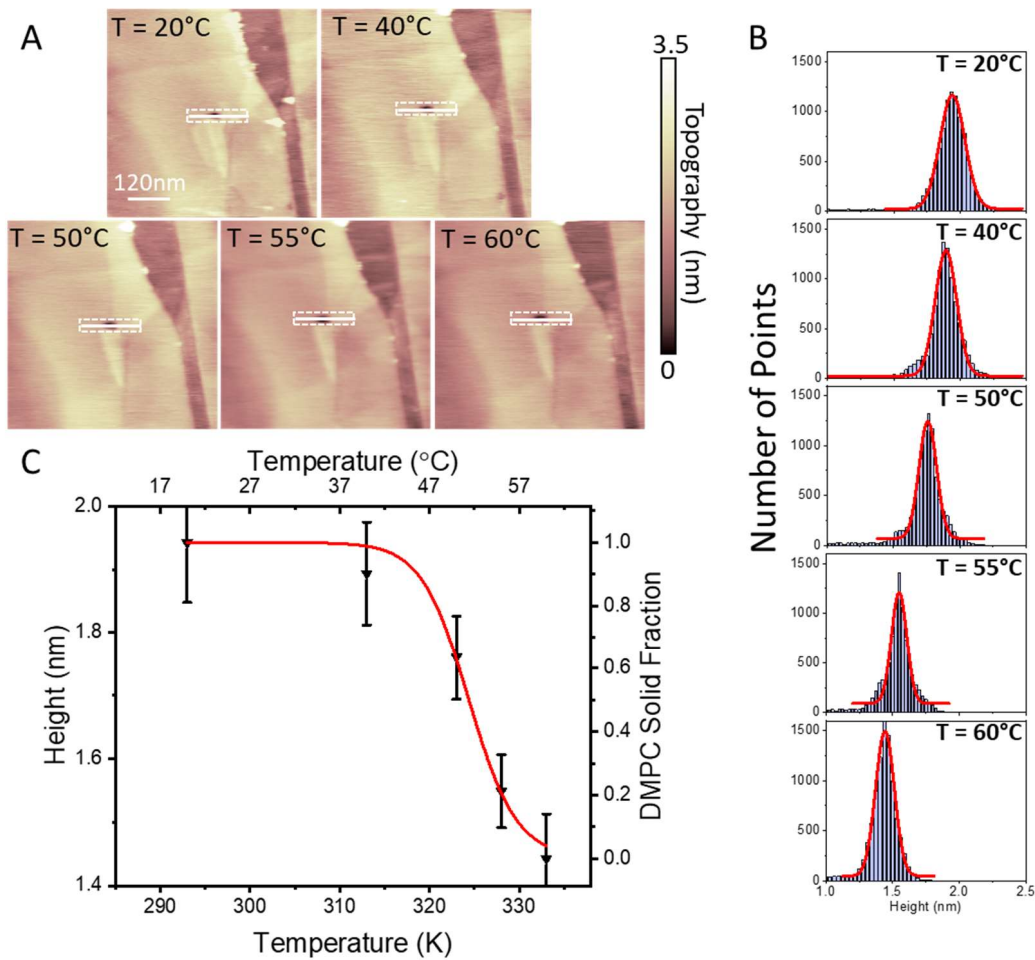

**Figure S3.** DMPC supported lipid monolayer formed on HOPG (0.1mg/ml DMPC/DI suspension). (a) A defect in the monolayer was used to measure the thickness of the monolayer with increasing temperature. (b) Height distributions show the decrease in monolayer thickness from 1.8 nm at 50 °C to 1.4 nm at 60 °C, indicating a phase transition from $S_o$ to $L_d$. (c) Temperature dependence of the membrane height (left axis) and fractional occupancy of the DMPC $S_o$ phase (right axis) - where the error bars represent the uncertainty in the height measurement.

**S3 Morphological features of DMPC and DLPC on HOPG and hBN**

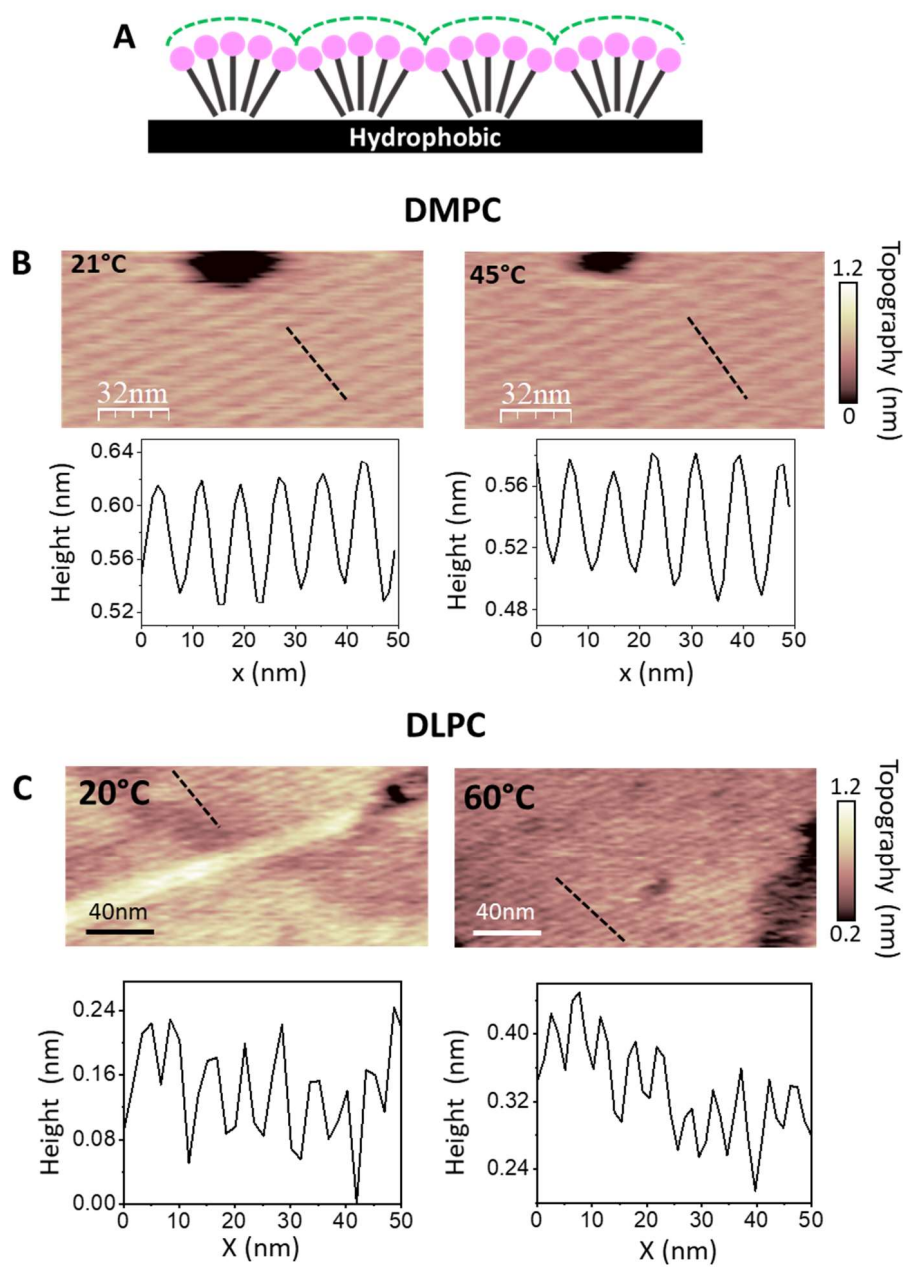

**Figure S4.** Morphology of zwitterionic lipid membranes formed on HOPG. The lipid monolayer shows stripe-like features, indicative of a hemimicellar conformation (a). Similar features were recorded for both DMPC (b) and DLPC (c). Importantly they were stable and did not change significantly with temperature.

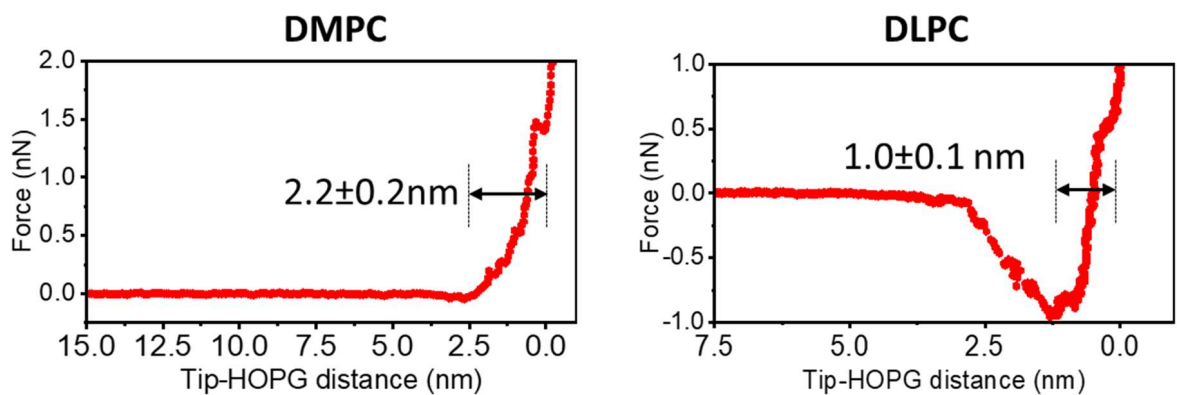

**Figure S5.** Force distance curves (FDC) obtained on lipid membranes formed on HOPG for both DMPC and DLPC. The breakthrough event into the lipid monolayer allows to determine the thickness of the lipid membrane.

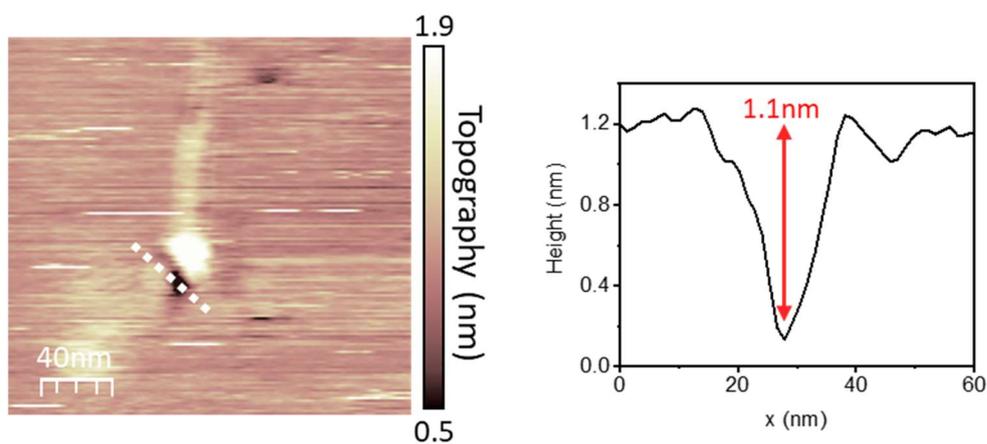

**Figure S6.** DLPC supported lipid membrane on HOPG. Topography map showing a defect in the lipid membrane from which a cross section was taken to determine the thickness of the membrane, which is consistent with the presence of a single monolayer adsorbed at the surface of HOPG.

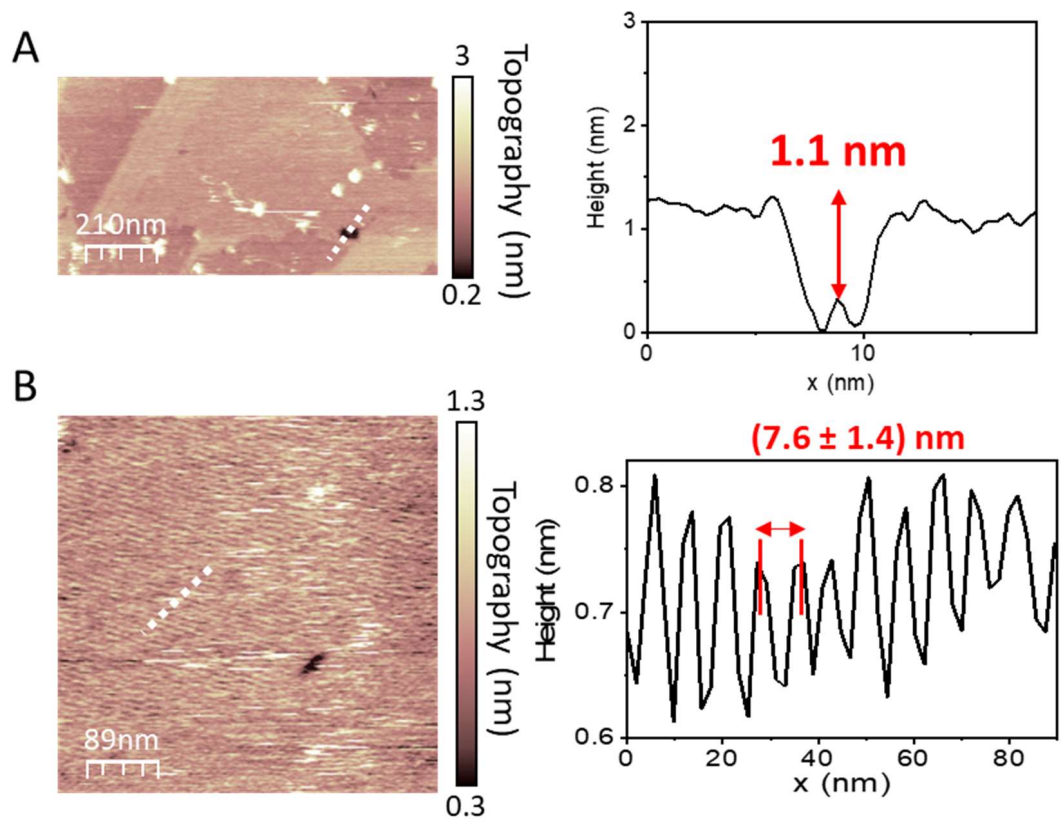

**Figure S7.** (a) Topography map showing a fully formed DLPC supported lipid membrane on h-BN at T = 7 °C (gel phase). A defect in the membrane was used to determine a membrane thickness of ~1.3 nm. The thickness agrees with that found on HOPG. (b) Topography map showing that like on HOPG, stripe-like structures in the membrane form. The periodicity of the ripples was measured to be around 7.5 nm.

**S4 Phase transition of DLPC monolayer on HOPG**

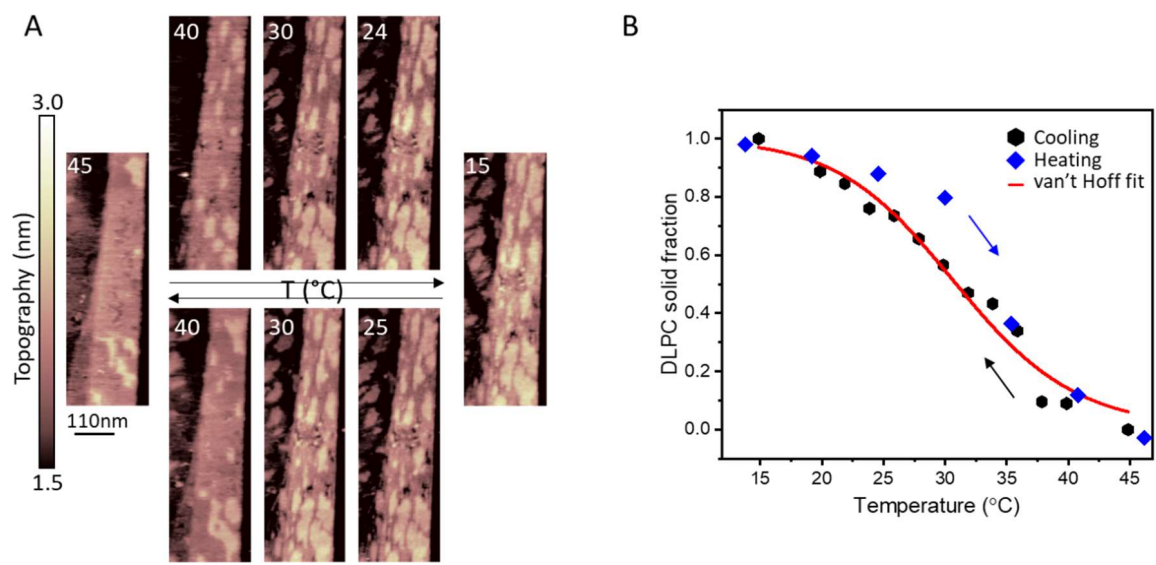

**Figure S8.** (a) AFM topography images as function of temperature (upon cooling and heating) of a DLPC membrane on HOPG. Taller domains appear with decreasing temperature (reversible process), indicating an equilibrium phase-transition. (b) The van't Hoff analysis was on the experimental data obtained during cooling, revealing a transition temperature of DLPC monolayer on HOPG of $T_m \approx 30.9$ °C.

**Supplementary Information References**